\documentclass[aps,prd,reprint,balancelastpage,nofootinbib,preprintnumbers,amsmath,amssymb,superscriptaddress,longbibliography]{revtex4-1}
\usepackage[
colorlinks=true,        % color link
citecolor=blue,         % cite color
linkcolor=blue,         % link color
urlcolor=blue           % url color
]{hyperref}             % create hyperlinks
\usepackage{bm}         % \bm{<text>} Bold math symbols
\usepackage{xcolor}     % textcolor
\usepackage{orcidlink}
\newcommand{\nc}{\newcommand*}

\nc{\Eq}[1]{Eq.~\eqref{#1}}     % equation
\nc{\Fig}[1]{Fig.~\ref{#1}}     % figure
\nc{\Table}[1]{Table~\ref{#1}}  % table
\nc{\Sec}[1]{Sec.~\ref{#1}}     % section
\def\({\left(}
\def\){\right)}
\def\[{\left[}
\def\]{\right]}
\def\e{\begin{equation}}
\def\q{\end{equation}}
\def\m{\begin{eqnarray}}
\def\n{\end{eqnarray}}

\begin{document}

\title{Pre-Big-Bang Cosmology Cannot Explain NANOGrav 15-year Signal}

%%%%%%%%%%%%%%%%%%%%%%%%%%%%%%%%%%%%%%%%%%%%%%%%%%%%%%%%%%%%%%%%%%%%%%%%%%%%%%%%%%%%%%%%%%%%%%%%
%%%%%%%%%%%%%%%%%%%%%%%%%%%%%%%%%%%%%%%%%%%%%%%%%%%%%%%%%%%%%%%%%
\author{Qin Tan\orcidlink{0000-0002-9496-6476}}
%\email{tanqin@hunnu.edu.cn}	
\affiliation{Department of Physics, Key Laboratory of Low Dimensional Quantum Structures and Quantum Control of Ministry of Education, Synergetic Innovation Center for Quantum Effects and Applications, Hunan Normal University, Changsha, 410081, Hunan, China}
\affiliation{Institute of Interdisciplinary Studies, Hunan Normal University, Changsha, Hunan 410081, China}

%%%%%%%%%%%%%%%%%%%%%%%%%%%%%%%%%%%%%%%%%%%%%%%%%%%%%%%%%%%%%%%%%
\author{You~Wu\orcidlink{0000-0002-9610-2284}}
%\email{Corresponding author: youwuphy@gmail.com}	
\email{youwuphy@gmail.com}
\affiliation{College of Mathematics and Physics, Hunan University of Arts and Science, Changde, 415000, China}

%%%%%%%%%%%%%%%%%%%%%%%%%%%%%%%%%%%%%%%%%%%%%%%%%%%%%%%%%%%%%%%%%
\author{Lang~Liu\orcidlink{0000-0002-0297-9633}}
%\email{Corresponding author: liulang@bnu.edu.cn}	
\email{ liulang@bnu.edu.cn}	
\affiliation{Faculty of Arts and Sciences, Beijing Normal University, Zhuhai 519087, China}

%%%%%%%%%%%%%%%%%%%%%%%%%%%%%%%%%%%%%%%%%%%%%%%%
\begin{abstract}
We investigate whether the Pre-Big Bang (PBB) scenario from string cosmology can explain the stochastic gravitational wave background signal reported in the NANOGrav 15-year dataset. Using Bayesian analysis techniques, we constrain the key parameters of the PBB model by comparing its theoretical predictions with the observed data. Our analysis yields $\beta = -0.12^{+0.06}_{-0.21}$ ($90\%$ credible interval) for the dilaton-dynamics parameter, which lies outside the theoretically allowed range $0 \lesssim \beta < 3$ with more than $5\sigma$ confidence. Additionally, model comparison strongly favors a simple power-law spectrum over the PBB scenario, with a Bayes factor of approximately $468$. These results demonstrate that the PBB scenario, in its current formulation, cannot adequately explain the NANOGrav observations, highlighting the need for either significant modifications to the model or alternative explanations for the observed signal.
\end{abstract}
\maketitle

%%%%%%%%%%%%%%%%%%%%%%%%%%%%%%%%%%%%%%%%%%%%%%%%%%%%%%%%%%%%%%%%%%%%%%%%%%%%%%%%%%%%%%%%%%%%%%%%
\section{Introduction}
Recently, the Pulsar timing array (PTA) community has made significant progress in the search for the stochastic gravitational wave background (SGWB). Several collaborations, including the Chinese PTA (CPTA)~\cite{Xu:2023wog}, the European PTA (EPTA) in collaboration with the Indian PTA (InPTA)~\cite{EPTA:2023sfo,Antoniadis:2023ott}, the Parkes PTA (PPTA)~\cite{Zic:2023gta,Reardon:2023gzh}, and the North American Nanohertz Observatory for GWs (NANOGrav)~\cite{NANOGrav:2023gor,NANOGrav:2023hde}, have presented compelling evidence for the presence of a stochastic signal in the frequency range of approximately $10^{-9}$ to $10^{-7}$ Hz. This collaborative effort has marked a milestone in the detection of SGWBs in the nanohertz frequency range, generating substantial interest in the scientific community due to its far-reaching implications. The opening of this new observational window has the potential to revolutionize our understanding of fundamental physics and shed light on the processes that shaped the early Universe~\cite{NANOGrav:2023hvm,EPTA:2023xxk}.

The SGWB is a diffuse background of gravitational waves (GW) that permeates the Universe. It is believed to be a superposition of GWs from a variety of unresolved sources, both astrophysical and cosmological in origin. Detecting and characterizing the SGWB is of utmost importance because it can provide unique insights into various processes that occurred in the early Universe and astrophysical phenomena~\cite{Caprini:2018mtu}. PTAs exploit the remarkable precision of millisecond pulsars, which act as stable celestial clocks, to detect the minute perturbations in the arrival times of radio pulses caused by passing GWs~\cite{Sazhin:1978myk}. The nanohertz frequency range probed by PTAs is particularly interesting because it corresponds to the characteristic frequencies of GWs from various cosmological sources, such as cosmic strings~\cite{Chen:2022azo,Kitajima:2023vre,Ellis:2023tsl,Wang:2023len,Ahmed:2023pjl,Antusch:2023zjk}, domain walls~\cite{Kitajima:2023cek,Blasi:2023sej,Babichev:2023pbf,Guo:2023hyp}, scalar-induced GWs~\cite{Inomata:2023zup,Chen:2019xse,Liu:2023ymk,Franciolini:2023pbf,HosseiniMansoori:2023mqh,Wang:2023ost,Jin:2023wri,Liu:2023pau,Zhao:2023joc,Yi:2023npi,Harigaya:2023pmw,Balaji:2023ehk,Yi:2023tdk,You:2023rmn,Liu:2023hpw,Choudhury:2023fwk,Choudhury:2023fjs,Domenech:2024rks,Chen:2024twp,Choudhury:2024dzw,Choudhury:2024aji,Chen:2024fir,Choudhury:2024kjj}, primordial GWs \cite{Vagnozzi:2020gtf,Benetti:2021uea,Vagnozzi:2023lwo,Jiang:2023gfe,Ye:2023tpz,Datta:2023vbs, Datta:2023xpr}, and phase transitions in the early Universe~\cite{Addazi:2023jvg,Athron:2023mer,Zu:2023olm,Jiang:2023qbm,Xiao:2023dbb,Abe:2023yrw,Gouttenoire:2023bqy,An:2023jxf,Chen:2023bms}. These exotic phenomena, predicted by various theories beyond the Standard Model of particle physics, could have left their imprints on the SGWB, providing a unique opportunity to test these theories and gain insights into the fundamental laws of nature. This makes PTAs an ideal instrument for investigating the SGWB that originated from the early Universe or new physics beyond the Standard Model~\cite{Sakharov:2021dim,Ashoorioon:2022raz,Chen:2023uiz,Madge:2023dxc,Dandoy:2023jot,Wu:2023pbt,Tan:2024esx}.

Another exciting possibility is that the detected signal is derived from string theory, a leading candidate for a unified theory of quantum gravity~\cite{Schwarz:1982jn,Kaplunovsky:1985yy}. String theory provides a consistent framework for unifying the fundamental interactions of nature~\cite{Witten:1995ex,Aharony:1999ti}, including gravity, and has made remarkable progress in addressing long-standing problems in theoretical physics, such as the Big Bang singularity~\cite{Borde:1993xh,Borde:2001nh} and the trans-Planckian problem~\cite{Martin:2000xs}. However, directly testing string theory predictions in laboratory experiments is extremely challenging, as the energy scales involved are far beyond the reach of current technology. Fortunately, the early Universe serves as a natural laboratory for testing string theory predictions, as the extreme conditions that prevailed in the primordial cosmos can leave observable imprints on the SGWB. In the context of string cosmology, the so-called Pre-Big Bang (PBB) scenario has emerged as a promising model that could produce SGWB~\cite{Gasperini:1992em,Gasperini:1996fu,Gasperini:2007vw,Fan:2008sh,Gasperini:2016gre,Li:2019jwh,Gasperini:2021mat,Jiang:2023qht,Ben-Dayan:2024aec}. This PBB phase would have generated a SGWB with a distinctive blue-tilted spectrum~\cite{Veneziano:1991ek}, where the amplitude of the GWs increases with frequency.

The PBB scenario appears to offer a natural framework for explaining the PTA observations. However, a detailed analysis is necessary to determine whether the specific predictions of PBB cosmology are compatible with the NANOGrav data. In this paper, we analyze the theoretical predictions of the PBB scenario for SGWB and compare these predictions with the NANOGrav 15-year dataset to constrain the parameter space of the PBB model. Our analysis reveals that the NANOGrav 15-year signal cannot be adequately explained by the PBB scenario, providing important constraints on string cosmology models.

The paper is organized as follows: Section~\ref{SGWB} reviews the theoretical framework of SGWB in PBB cosmology. Section~\ref{data} presents our analysis methodology and results using the NANOGrav dataset. We discuss our conclusions and their implications in Section~\ref{conclusion}.

%%%%%%%%%%%%%%%%%%%%%%%%%%%%%%%%%%%%%%%%%%%%%%%%%%%%%%%%%%%%%%%%%%%%%%%%%%%%%%%%%%%%%%%%%%%%%%%%
\section{ \label{SGWB} SGWB from PBB Scenario}

Based on the scale-factor duality of the string cosmological equation~\cite{Veneziano:1991ek}, the so-called PBB scenario can produce a SGWB with a blue-tilted (i.e. increasing with frequency) spectrum. The present-day spectral energy density of the SGWB within our cosmic horizon can be expressed as:
\begin{equation}
\Omega_{\textsc{gw}}(k,\tau_{0}) = \frac{1}{\rho_{\rm crit}(\tau_{0})} \frac{d \rho_{\textsc{gw}}}{d \ln k},\label{eq:Omega1}
\end{equation}
where $\tau_{0}$ denotes the current conformal time, and $\rho_{\text{crit}}=3M_{\text{Pl}}^{2}H^{2}$ represents the critical energy density. We focus on the contribution to the SGWB from the cosmological amplification of metric tensor fluctuations. The energy density of each mode $k$ is 
\begin{equation}
d\rho(\tau_{0})=2k\langle n_{k}(\tau_{0})\rangle\frac{d^{3}k}{8\pi^{3}}=\frac{k^{4}}{\pi^{2}}\langle n_{k}(\tau_{0})\rangle\ln k,\label{eq:energy density1}
\end{equation}
where $\langle n_{k}(\tau_{0})\rangle$ represents the graviton number density at $\tau_{0}$. To obtain $\langle n_{k}(\tau_{0})\rangle$, we solve the evolution equation for the tensor perturbation mode $h_k$~\cite{Gasperini:2016gre}:
\begin{equation}
v_k'' + \left( k^2 - \frac{\xi''}{\xi} \right) v_k=0,\label{eq:evolution equation}
\end{equation}
where prime denotes the derivative with respect to conformal time $\tau$, $v_{k}=\xi h_{k}$, and $\xi(\tau)$ is the ``pump field" that determines the dynamics of the fluctuation $h_{k}$. In the chosen model, the background is approximated as a sequence of five cosmic phases, with the pump field $\xi$ exhibiting power-law behavior in each phase: $\xi=(M_{\text{Pl}}/\sqrt{2})|\tau/\tau_{1}|^{\alpha}$, where $\tau_{1}$ marks the end of the string phase. The solution $h_k$ of Eq.\eqref{eq:evolution equation} can be expressed using Hankel functions $H^{(1)}_{\nu}$ and $H^{(2)}_{\nu}$ as
\begin{eqnarray}
h_k(\tau) &=& \left(\frac{2 \tau_1}{M_\text{Pl}^2}\right)^{\frac12} \left | \frac{\tau}{\tau_1} \right|^\nu \bigg[ A_+(k) H_\nu^{(2)}(k\tau)\nonumber\\
&&+ A_-(k) H_\nu^{(1)}(k\tau)\bigg],
\label{eq:hksolution}
\end{eqnarray}
where $\nu=\frac{1}{2}-\alpha$, and $A_{\pm}$ are coefficients determined by the continuity of $h_k$ and $h'_k$ in each phase and the initial condition $v_{k}=(1/\sqrt{2k})\exp(-ik\tau)$ for $\tau\rightarrow-\infty$. The graviton number density $\langle n_{k}(\tau_{0})\rangle$ is then given by
\begin{equation}
\langle n_{k}(\tau_{0})\rangle=\frac{4}{\pi}|A_{-}(k)|{\tau=\tau{0}}.\label{eq:number density}
\end{equation}
Combining this equation with Eqs.~\eqref{eq:Omega1} and~\eqref{eq:energy density1} yields the SGWB produced by the PBB scenario.

The model considered here is divided into five phases separated by four transitions (at $\tau_{i}:\tau_1,\tau_\sigma,\tau_d,\tau_s$), corresponding to the end of the string phase, the onset of an axion-dominated dust phase, the beginning of post-Big-Bang evolution, and the transition from a low-energy initial stage to a possible late attractor, respectively. The specific form of $\xi$ is given by~\cite{Gasperini:2016gre}
\begin{equation}
	\xi \sim \left\{
	\begin{aligned}
 \hspace{-.1cm}
  &\frac{M_\text{Pl}}{\sqrt{2}}(-\tau)^{1/2},~~~~~ \tau< -\tau_{s} \\ 
  &\frac{M_\text{Pl}}{\sqrt{2}}(-\tau)^{\beta-1},~~~  -
\tau_{s}<\tau< -\tau_{1} \\ 
  &\frac{M_\text{Pl}}{\sqrt{2}}\tau,~~~~~~~~~~~~~ -\tau_{1}<\tau< \tau_{\sigma} \\ 
  &\frac{M_\text{Pl}}{\sqrt{2}}\tau^{2},~~~~~~~~~~~~  \tau_{\sigma}<\tau< \tau_{d} \\ 
   &\frac{M_\text{Pl}}{\sqrt{2}}\tau,~~~~~~~~~~~~~ \tau_{d}<\tau< \tau_{\text{eq}}
	\end{aligned}\right.
	\label{piecepbb}
\end{equation}
where $\beta$ is a parameter describing the dynamics of internal dimensions and the high-energy growth of the dilaton. 
% \textcolor{red}{The definition of $\beta$ is~\cite{Conzinu:2024cwl}
% \begin{equation}
% \beta = \frac{d \log g_s}{d \log a},\label{cons_beta}
% \end{equation}
% where $g_s$ is the effective four-dimensional string coupling and $a$ is the scale factor. For the string coupling to grow~\cite{Gasperini:2023tus,Conzinu:2023fth,Modesto:2022asj}, we need $\beta \gtrsim 0$. On the other hand, the growth of the string coupling parameter cannot be too fast. If it grows too quickly, it will cause the comoving amplitude of fluctuations to grow rapidly, breaking the validity of the linear approximation, which will lead to quantum instability of the background~\cite{Kawai:1998ab}. Therefore, we further require that at the end of the string phase, $g_s \sim \mathcal{O}(1)$. Consequently, combining the requirements from both aspects, $\beta$ is constrained to $0 \lesssim \beta \lesssim 3$.}

The dilaton-dynamics parameter $\beta$ plays a pivotal role in the PBB scenario, as it describes the dynamics of the internal dimensions and the high-energy growth of the dilaton field. The theoretically allowed range for $\beta$ is constrained to $0 \lesssim \beta < 3$, a restriction that arises from both stability requirements~\cite{Kawai:1998ab} and the necessity of a growing string coupling for a smooth bounce transition~\cite{Gasperini:2023tus,Conzinu:2023fth,Modesto:2022asj}.

Firstly, the lower bound of $\beta \gtrsim 0$ is necessary to ensure the growth of the string coupling during the pre-big bang phase. In the PBB scenario, the string coupling is inversely proportional to the exponential of the dilaton field. As demonstrated by Refs.~\cite{Gasperini:2023tus,Conzinu:2023fth,Modesto:2022asj},  a positive $\beta$ leads to a growing string coupling, which is essential for a smooth transition to the post-big bang era. Negative values of $\beta$ would result in a decreasing string coupling, preventing a graceful exit from the PBB phase.

Secondly, the upper limit of $\beta<3$ is imposed by the requirement of background stability. As shown by Ref.~\cite{Kawai:1998ab}, values of $\beta \geq 3$ lead to an instability in the one-loop superstring cosmology, rendering the scenario unviable. Therefore, to ensure a stable cosmological evolution, $\beta$ must be strictly less than 3.

These theoretical constraints on $\beta$ are fundamental to the PBB scenario and are rooted in the underlying string theory framework. Violating these bounds would lead to instabilities or inconsistencies in the cosmological model, undermining its viability. Therefore, any observational evidence that points to values of $\beta$ outside the range $0 \lesssim \beta < 3$ would pose a significant challenge to the PBB scenario.

Based on the specific forms of the pump field, the energy density fraction spectrum of the SGWB can be represented as~\cite{Gasperini:2016gre,Ben-Dayan:2024aec}
\begin{equation}
 \Omega_{\text{GW}}(f) = \left\{
	\begin{aligned}
		\hspace{-.1cm}
  &\Omega_{\rm PBB}\exp{[-(f-f_1)/f_1]},~~~~   f >f_1 \\ 
   &\Omega_{\rm PBB}\left(\dfrac{f}{f_1}\right)^{ \beta_1},~~~~~~~~~~~~~~~~ f_\sigma \lesssim f \lesssim f_1 \\ 
 &\Omega_{\text{gw}}(f_1)\left(\dfrac{f_\sigma}{f_1}\right)^{\beta_1}
\left(\dfrac{f}{f_\sigma}\right)^{\beta_2}, ~ f_d \lesssim  f \lesssim f_\sigma \\ 
 &\Omega_{\text{gw}}(f_\sigma)\left(\dfrac{f_d}{f_\sigma}\right)^{\beta_2}\left(\dfrac{f}{f_d}\right)^{\beta_1},~ f_s \lesssim f \lesssim f_d\\ 
 &\Omega_{\text{gw}}(f_d)\left(\dfrac{f_s}{f_d}\right)^{\beta_1}\left(\dfrac{f}{f_s}\right)^{3},~~~ f \lesssim f_s
	\end{aligned}\right.
	\label{eq:spectrum main}
\end{equation}
where $f_{i}=1/(2\pi \tau_{i})$, $\beta_1=3- |3-2 \beta|$, and $\beta_2=1- |3-2 \beta|$. The dimensionless amplitude $\Omega_{\text{PBB}}$ is given by
\begin{equation}
\Omega_{\text{PBB}}=\Omega_{\text{r}0}\left(\frac{H_1}{M_{\text{Pl}}}\right)^{2} \left(\frac{f_{d}}{f_{\sigma}}\right)^{2},
\end{equation}
with $\Omega_{\text{r}0}\approx4.15\times10^{-5}h^{-2}$ being the critical fraction of the current radiant energy density. For convenience, we define three parameters~\cite{Ben-Dayan:2024aec}
\begin{equation}
z_{s}=\frac{\tau_{s}}{\tau_{1}}=\frac{f_{1}}{f_{s}},~~~~~~
z_{\sigma}=\frac{\tau_{\sigma}}{\tau_{1}}=\frac{f_{1}}{f_{\sigma}},~~~~~~
z_{d}=\frac{\tau_{d}}{\tau_{1}}=\frac{f_{1}}{f_{d}}.
\end{equation}
The frequencies $f_1$ and the corresponding curvature scales $H_1=H(\tau_{1})$ can be expressed using these newly defined parameters as
\begin{equation}
f_{1}=\frac{3.9\times10^{11}}{2\pi}\left(\frac{H_{1}}{M_{\text{Pl}}}\right)^{1/2}\left(\frac{z_{\sigma}}{z_{d}}\right)^{1/2}\mathrm{Hz},\label{eq:f1}
\end{equation}
and
 \begin{eqnarray}
   \log_{10}\left(\frac{H_{1}}{M_{\text{Pl}}}\right)&=&\frac{2}{5 - n_{\rm s}} \Bigg\{ \log_{10} \left[\frac{4.2 \pi^2}{T^2(H_{1})}\right] -9 \nonumber\\
   &&+ (1-n_{\rm s})(\log_{10}  1.5 - 27)\nonumber\\
	&&  + (1-n_{\rm s} -2 \beta) \log_{10}  z_s \nonumber\\
 &&+\frac{n_{\rm s}-1}{2} \left(\log_{10}  z_\sigma - \log_{10}  z_d \right) \Bigg\}\label{eq:H1}
  \end{eqnarray}
with
 \m
  n_\text{s}&=&0.9649\pm0.0042,\label{eq:ns}\\
  T(H_{1})&\approx& 0.13\left(\frac{H_{1}}{M_{\text{Pl}}}\right)^{1/6}z_{d}^{1/4}z_{\sigma}^{-\frac{7}{12}}\nonumber\\
  && + 0.25\left(\frac{H_{1}}{M_{\text{Pl}}}\right)^{-1/6}z_{d}^{-1/4}z_{\sigma}^{\frac{7}{12}}-0.01.\label{eq:T}
 \n

Thus, the SGWB spectrum~\eqref{eq:spectrum main} is determined by four undetermined parameters: $\beta, z_{s}, z_{\sigma}$, and $z_{d}$. We will use  NANOGra data to constrain these parameters, providing insights into the search for observational features of string theory.

%%%%%%%%%%%%%%%%%%%%%%%%%%%%%%%%%%%%%%%%%%%%%%%%%%%%%%%%%%%%%%%%%%%%%%%%%%%%%%%%%%%%%%%%%%%%%%%%
\section{\label{data}Data analysis and results}

We analyze the PTA signal detected in the NANOGrav 15-year dataset to test predictions of the PBB cosmological model. This investigation leverages timing observations from an array of $67$ millisecond pulsars monitored over a baseline of $T_{\mathrm{obs}}=16.03$ years~\cite{NANOGrav:2023hde,NANOGrav:2023gor}.

The analysis framework centers on the characteristic Hellings-Downs spatial correlation pattern~\cite{Hellings:1983fr}, which provides a crucial discriminator between genuine GW signals and various noise sources. The power spectral density $S(f)$ of the GW strain is reconstructed from the timing residual cross-correlations through:
\begin{equation}
S(f) = d(f)^2 T_{\mathrm{obs}},
\end{equation}
where $d(f)$ represents the measured time delay at frequency $f$.

The NANOGrav analysis employs $14$ frequency bins spanning $2.0 \times 10^{-9}$ Hz to $2.8 \times 10^{-8}$ Hz. This frequency range is particularly sensitive to cosmological signatures from the PBB epoch. We compute the dimensionless energy density spectrum $\hat{\Omega}_{\mathrm{GW}}(f)$ using:
\begin{equation}\label{hatOgw}
\hat{\Omega}_{\mathrm{GW}}(f) = \frac{8\pi^4}{H_0^2} T_{\mathrm{obs}} f^5 d^2(f),
\end{equation}
where we adopt the Planck value $H_0 = 67.4\, \mathrm{km}\,\mathrm{s}^{-1}\,\mathrm{Mpc}^{-1}$~\cite{Planck:2018vyg}. The characteristic strain spectrum $h_c(f)$ is related to the power spectrum through:
\begin{equation}
h_c^2(f) = 12 \pi^2 f^3 S(f).
\end{equation}

To constrain the PBB model parameters $\Lambda = \{\beta, z_s, z_d, z_\sigma\}$, we construct a likelihood function incorporating kernel density estimates (KDEs) of the posterior distributions for each frequency bin~\cite{Moore:2021ibq,Lamb:2023jls,Liu:2023ymk,Wu:2023hsa,Jin:2023wri,Liu:2023pau}:
\begin{equation}
\ln \mathcal{L}(\Lambda) = \sum_{i=1}^{14} \ln \mathcal{L}_i(\Omega_{\mathrm{GW}}(f_i, \Lambda)).
\end{equation}
For each frequency bin $f_i$, we construct a likelihood function based on the posterior distribution of $\hat{\Omega}_{\mathrm{GW}}(f_i)$ obtained from the NANOGrav analysis as described in \Eq{hatOgw}.
We implement this using KDE to approximate the posterior distribution at each frequency:
\begin{equation}
    \mathcal{L}_i(\Omega_\mathrm{GW}(f_i,\Lambda)) = \frac{1}{N h} \sum_{j=1}^{N} K\left(\frac{\Omega_\mathrm{GW}(f_i,\Lambda) - \hat{\Omega}_\mathrm{GW}^{(j)}(f_i)}{h}\right),
\end{equation}
where $K(x) = \frac{1}{\sqrt{2\pi}}e^{-x^2/2}$ is the Gaussian kernel, $\hat{\Omega}_\mathrm{GW}^{(j)}(f_i)$ represents the $j$-th posterior sample from the NANOGrav analysis at frequency $f_i$, $N$ is the total number of posterior samples, and $h$ is the bandwidth parameter determined using Scott's rule as implemented in the \texttt{Scipy}~\cite{2020SciPy-NMeth} package. This approach is valid because NANOGrav employed uniform priors in their analysis when obtaining the measured time delay $d(f)$, allowing us to use the posterior samples directly to construct our likelihood function.

%%%%%%%%%%%%%%%%%%%%%%%%%%%%%%%%%%%%%%%%%%%%%%%%%%%%%%%%%%%%%%%%%
\begin{table}
    \centering
	\begin{tabular}{cccc}
		\hline
		Parameter & Prior & Result \\ 
  \hline \\[-2\medskipamount]   
  $\beta$ & Uniform$[-1, 3]$ & $-0.12^{+0.06}_{-0.21}$\\[1pt]
  $\log_{10}z_s$ & Uniform$[5, 20]$ & $15.5^{+1.3}_{-2.2}$\\[1pt]
  $\log_{10}z_d$ & Uniform$[-2, 20]$ & $9.9^{+4.4}_{-7.0}$\\[1pt]
  $\log_{10}z_\sigma$ & Uniform$[-2, 20]$ & $6.3^{+6.4}_{-4.9}$\\[1pt]
  \hline
	\end{tabular}
	\caption{\label{tab:prior}Prior distributions and posterior constraints for the PBB model parameters. The posterior values are reported as medians with symmetric $90\%$ credible intervals.}
\end{table}
%%%%%%%%%%%%%%%%%%%%%%%%%%%%%%%%%%%%%%%%%%%%%%%%%%%%%%%%%%%%%%%%%%%%%%%%%%%%%%%%%%%%%%%%%%%%%%%%

%%%%%%%%%%%%%%%%%%%%%%%%%%%%%%%%%%%%%%%%%%%%%%%%%%%%%%%%%%%%%%%%%
\begin{figure}[tbp!]
	\centering
 \includegraphics[width=0.5\textwidth]{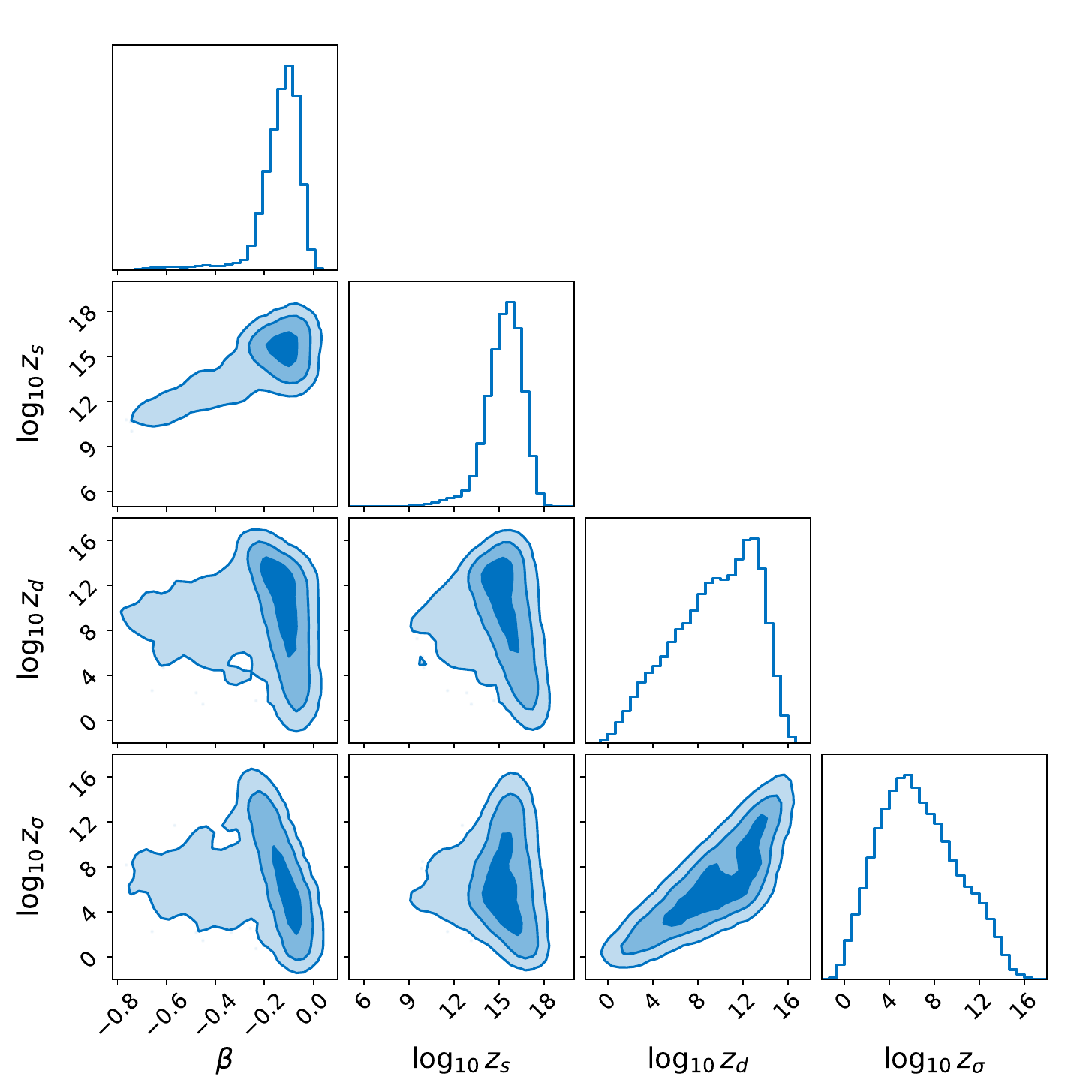}
	\caption{\label{posts} Joint and marginal posterior distributions for the PBB model parameters $\Lambda = \{\beta, z_s, z_d, z_\sigma\}$ from the NANOGrav 15-year dataset. Diagonal panels show marginalized posteriors, while off-diagonal panels display joint distributions with $1 \sigma$, $2 \sigma$, and $3 \sigma$ credible regions.}
\end{figure}

We explore the parameter space using nested sampling implemented through \texttt{dynesty}~\cite{Speagle:2019ivv} within the \texttt{Bilby} framework~\cite{Ashton:2018jfp}. The prior distributions for all model parameters are summarized in Table~\ref{tab:prior}. Following the theoretical constraints outlined in~\cite{Ben-Dayan:2024aec}, the parameter space must satisfy:
\begin{align}
&1 \lesssim z_\sigma<z_d<z_s,\\
&\log _{10}\left(\frac{H_1}{M_{\mathrm{Pl}}}\right)+\frac{3}{2} \log _{10} z_d-\frac{7}{2} \log _{10} z_\sigma \leqslant 0,\\
&\log _{10}\left(\frac{H_1}{M_{\mathrm{Pl}}}\right)-3 \log _{10} z_d+\log _{10} z_\sigma>-42-\log _{10} 4,\\
&\log _{10} z_s<26-\log _{10} 9+\frac{1}{2} \log _{10}\left(\frac{H_1}{M_{\mathrm{Pl}}}\right)\nonumber\\ 
&\qquad \qquad \qquad +\frac{1}{2}\left(\log _{10} z_\sigma-\log _{10} z_d\right).
\end{align}

%%%%%%%%%%%%%%%%%%%%%%%%%%%%%%%%%%%%%%%%%%%%%%%%%%%%%%%%%%%%%%%%%
\begin{figure}[tbp!]
\centering
\includegraphics[width=0.5\textwidth]{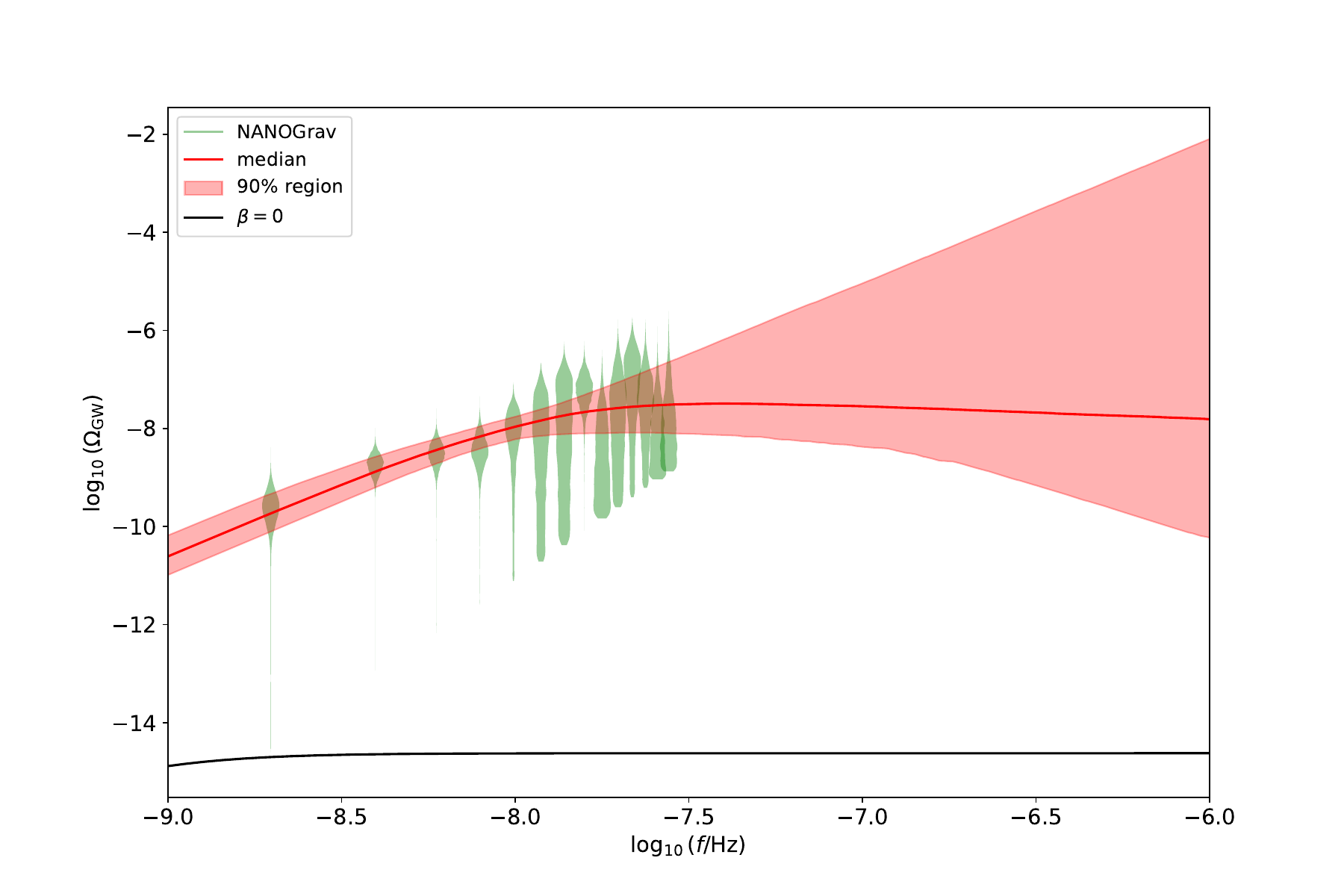}
\caption{\label{ogw} Energy density spectrum of the SGWB predicted by the PBB model. Green violin plots show the posterior predictive distribution from the NANOGrav 15-year dataset, with the red shaded region indicating the $90\%$ credible interval obtained from NANOGrav data. The black line shows the theoretical spectrum for $\beta=0$, demonstrating the incompatibility between PBB predictions and the observed data.}
\end{figure}

%%%%%%%%%%%%%%%%%%%%%%%%%%%%%%%%%%%%%%%%%%%%%%%%%%%%%%%%%%%%%%%%%
\begin{figure}[tbp!]
\centering
\includegraphics[width=0.5\textwidth]{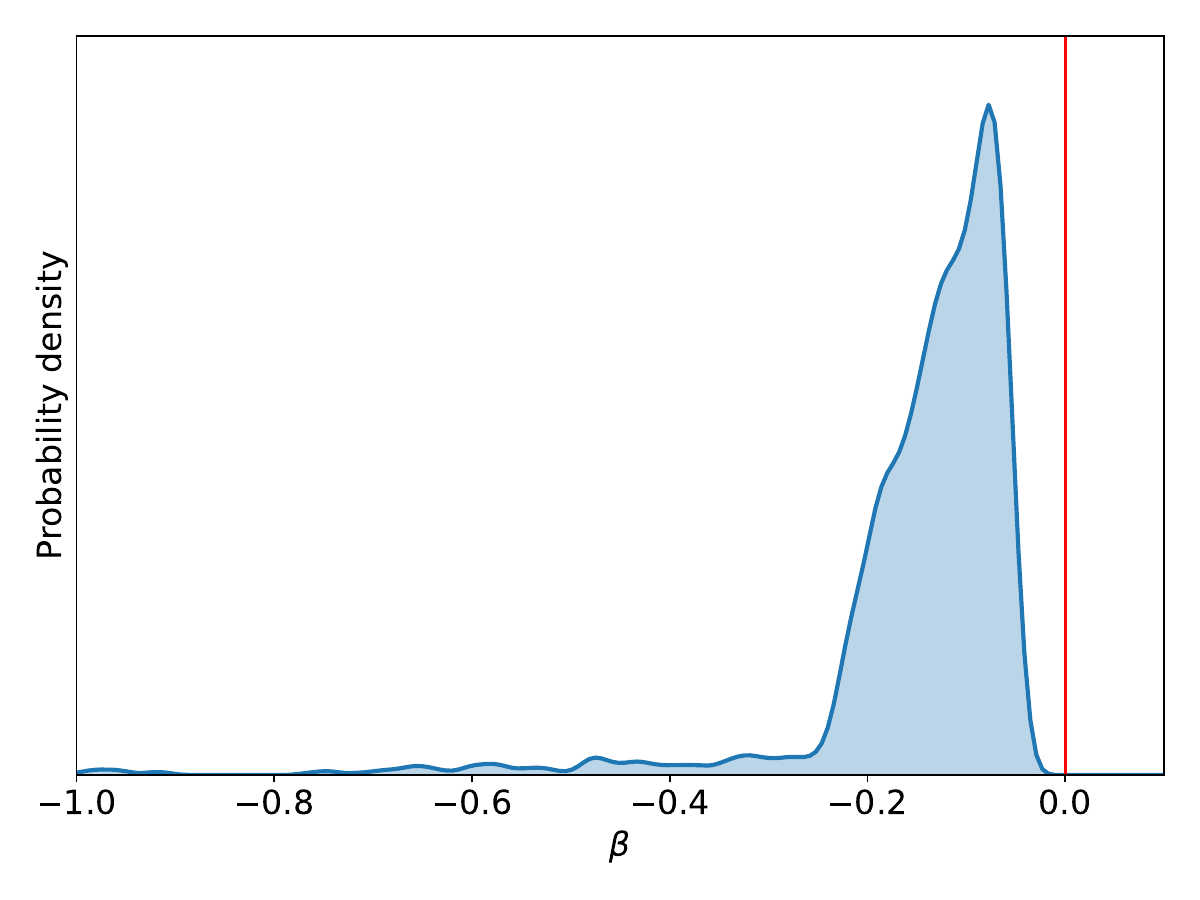}
\caption{\label{post_beta}Marginalized posterior distribution for the dilaton-dynamics parameter $\beta$. The theoretically allowed range $0 \lesssim \beta < 3$ lies entirely outside the observed distribution, indicating a strong tension between PBB predictions and NANOGrav data.}
\end{figure}
  
Our analysis yields the following parameter constraints (90\% credible intervals): $\beta = -0.12^{+0.06}_{-0.21}$, $\log_{10}z_s = 15.5^{+1.3}_{-2.2}$, $\log_{10}z_d = 9.9^{+4.4}_{-7.0}$, and $\log_{10}z_\sigma =6.3^{+6.4}_{-4.9}$ (see Table~\ref{tab:prior} and Figure~\ref{posts}). 
{As evident in \Fig{posts}, the posterior distributions exhibit extended tails, particularly for $\beta$ and $\log_{10} z_s$. This feature arises from the positive correlation between these parameters, reflecting how the dilaton dynamics parameter $\beta$ and the transition scale $z_s$ jointly influence the shape of the GW spectrum needed to match the NANOGrav observations.} 
The posterior predictive distribution for the GW energy density spectrum $\Omega_{\mathrm{GW}}(f)$ is shown in Figure~\ref{ogw}.
A significant tension emerges from our analysis: the inferred value of $\beta$ is smaller than $-0.06$ at the $5\sigma$ confidence level (see also Figure~\ref{post_beta}). This result directly conflicts with the theoretical constraint $0 \lesssim \beta < 3$, which has been discussed in Sec.~\ref{SGWB}. 

In addition to parameter estimation, we perform Bayesian model comparison to evaluate whether the PBB scenario or a simpler power-law model better explains the NANOGrav data. The Bayes factor, which quantifies the relative evidence between two models, is defined as:
\begin{equation} 
\mathcal{B}_{12} = \frac{\mathcal{Z}_1}{\mathcal{Z}_2},
\end{equation}
where $\mathcal{Z}_i$ represents the evidence (marginal likelihood) for model $i$, calculated as:
\begin{equation} 
\mathcal{Z}_i = \int \mathcal{L}(\mathbf{d}|\theta_i, \mathcal{M}_i) \pi(\theta_i|\mathcal{M}_i)\, d\theta_i.
\end{equation}
Here, $\mathcal{L}(\mathbf{d}|\theta_i, \mathcal{M}_i)$ is the likelihood of the data $\mathbf{d}$ given the parameters $\theta_i$ of model $\mathcal{M}_i$, and $\pi(\theta_i|\mathcal{M}_i)$ is the prior distribution for those parameters.
For our comparison, we consider two models: 
\begin{itemize}
\item The PBB model with four parameters $\Lambda =\{\beta, z_s, z_d, z_\sigma\}$ as described in Section~\ref{SGWB}, which produces a complex spectral shape given by \Eq{eq:spectrum main}. 
\item A simple power-law model characterized by two parameters: an amplitude $A$ and a spectral index $\gamma$, with energy density spectrum: 
\begin{equation} 
\Omega_{\mathrm{GW}}(f) = A \left(\frac{f}{f_{\mathrm{yr}}}\right)^{\gamma},
\end{equation} 
where $f_{\mathrm{yr}} = 1$ yr$^{-1}$. For this model, we adopt log-uniform priors on the amplitude $A \in [10^{-18}, 10^{-11}]$ and uniform priors on the spectral index $\gamma \in [0, 8]$.
\end{itemize}
We compute the evidences for both models using the nested sampling algorithm implemented in \texttt{dynesty}, which provides robust estimates of the Bayesian evidence along with the parameter posterior distributions. A Bayes factor $\mathcal{B}_{12} > 1$ indicates that model 1 is favored over model 2, with values $\mathcal{B}_{12} > 100$ conventionally interpreted as strong evidence according to the Jeffreys' scale~\cite{Kass:1995loi}.

The Bayesian model comparison strongly favors a simple power-law spectrum (typically attributed to supermassive black hole binaries~\cite{Sesana:2008mz,Chen:2018rzo,Bi:2023tib}) over the PBB scenario, with a Bayes factor of $468$, indicating strong evidence favoring the simple power-law model over the PBB model~\cite{BF}. These results indicate that the PBB model, in its current formulation, cannot adequately explain the NANOGrav observations.

%%%%%%%%%%%%%%%%%%%%%%%%%%%%%%%%%%%%%%%%%%%%%%%%%%%%%%%%%%%%%%%%%%%%%%%%%%%%%%%%%%%%%%%%%%%%%%%%
\section{\label{conclusion}Conclusion}

In this work, we have conducted a comprehensive analysis of the PBB cosmological scenario as a potential explanation for the PTA signal detected in the NANOGrav 15-year dataset. Our investigation yields several significant findings that pose fundamental challenges to the PBB interpretation of the data.

First and foremost, our analysis reveals that the parameter $\beta$, which describes the dynamics of internal dimensions and dilaton evolution, is constrained to be $\beta = -0.12^{+0.06}_{-0.21}$ at the 90\% confidence level. This result is problematic because it lies outside the theoretically allowed range of $0 \lesssim \beta < 3$, which is required for both stability considerations and the necessity of growing string coupling for a smooth bounce transition. The violation of this theoretical bound is significant at the $5\sigma$ level, indicating a fundamental incompatibility between the PBB predictions and the observed data.

Furthermore, our Bayesian model comparison strongly disfavors the PBB scenario when compared to a simple power-law model, with a Bayes factor of approximately $468$ in favor of the latter. This suggests that the complexity introduced by the PBB model is not justified by its ability to explain the data, particularly when compared to simpler alternatives such as the SGWB from supermassive black hole binaries.

The constraints we have derived on the other model parameters ($z_s$, $z_d$, $z_\sigma$) provide additional insights into the viability of the PBB scenario. While these parameters are well-constrained by the data, the overall framework fails to provide a satisfactory explanation for the observed signal characteristics.

These results have important implications for string cosmology and early universe physics. They suggest that either significant modifications to the current PBB framework are necessary, or alternative theoretical approaches should be pursued to explain the NANOGrav observations. Our findings emphasize the power of GW observations as a tool for testing fundamental theories of the early universe and highlight the challenges faced by string-inspired cosmological models in explaining observational data.

Future observations with improved sensitivity and extended frequency coverage will be crucial in further testing these conclusions and potentially identifying the true origin of the signal detected by NANOGrav.

Note added: A similar model reaching the same qualitative conclusions as ours,  as well as a modified model claimed to fit the NANOGrav data, were presented in  Ref.~\cite{Conzinu:2024cwl}, which appeared briefly after the submission of this work.

%%%%%%%%%%%%%%%%%%%%%%%%%%%%%%%%%%%%%%%%%%%%%%%%%%%%%%%%%%%%%%%%%
\begin{acknowledgments}
We thank the authors of \cite{Conzinu:2024cwl} for pointing a few typos and errors in the first arXiv version, which are now corrected on the second arxiv version. These do not affect our results at
qualitative level. 
QT is supported by the National Natural Science Foundation of China (Grants No.~12405055 and No.~12347111), the China Postdoctoral Science Foundation (Grant No.~2023M741148), the Postdoctoral Fellowship Program of CPSF (Grant No. GZC20240458), and the innovative research group of Hunan Province under Grant No.~2024JJ1006.
YW is supported by the National Natural Science Foundation of China under Grant No.~12405057.
LL is supported by the National Natural Science Foundation of China Grant under Grant No.~12433001.  
\end{acknowledgments}

\bibliography{ref}
\end{document}